\documentclass[preprintnumbers,prd,twocolumn,showpacs,floatfix,preprintnumbers,letterpaper,nofootinbib,superscriptaddress]{revtex4}
\usepackage{epsfig}
\usepackage{bm}
\usepackage{amsfonts}
\usepackage{subfigure}
\usepackage{amsmath,amssymb,latexsym}
\usepackage{epstopdf}
\usepackage{tabulary,tabularx}
\usepackage{pst-grad}
\newcommand{\p}{\partial}
\newcommand{\eq}{\begin{equation}}
\newcommand{\eqe}{\end{equation}}
\newcommand{\nn}{\nonumber}

\newcommand{\eqa}{\begin{eqnarray}}
\newcommand{\eqae}{\end{eqnarray}}

\newcommand{\Om}{\Omega_{ DE}}

\begin{document}

\title{A Quintom scenario with mixed kinetic terms}

\author{Emmanuel N. Saridakis}
 \email{msaridak@phys.uoa.gr}
 \affiliation{College of Mathematics and Physics, Chongqing University of Posts and
Telecommunications,\\ Chongqing, 400065, P.R. China }

\author{Joel M. Weller}
 \email{app07jmw@shef.ac.uk}
\affiliation{ Department of Applied Mathematics, University of
Sheffield, Hounsfield Road, Sheffield S3 7RH, United Kingdom }

\begin{abstract}

We examine an extension of the quintom scenario of dark energy, in
which a canonical scalar field and a phantom field are coupled through a
kinetic interaction. We perform a phase space analysis and show
that the kinetic coupling gives rise to novel
cosmological behaviour. In particular, we obtain both
quintessence-like and phantom-like late-time solutions, as well as
solutions that cross the phantom divide during the evolution of the Universe.
\end{abstract}

 \pacs{98.80.-k, 95.36.+x}

\maketitle

\section{Introduction}

According to independent observational data, including measurements of
 type Ia supernovae {\cite{obs1}}, the Wilkinson Microwave Anisotropy Probe {\cite{obs2}},
  the Sloan Digital Sky Survey {\cite{obs3}}, and X-ray observations
{\cite{obs4}}, the Universe is undergoing a phase of accelerated
expansion. Although the cosmological constant seems to be a simple
and economic way to explain this behaviour \cite{cosmcont}, the
extreme degree of fine-tuning associated with this means that it
is attractive to consider other dark energy candidates. The
dynamical nature of dark energy, at least at an effective level,
can originate from various fields, such as a canonical scalar
field (quintessence) \cite{quint} or a phantom field  \cite{phant}
(i.e. a scalar field with negative kinetic terms). Another
possibility of considerable interest is  the quintom scenario
\cite{Guo:2004fq,quintom, Cai:2009zp}, in which the acceleration
is driven by a combination of quintessence and phantom fields. The
quintom paradigm has the advantage of allowing the dark energy
equation-of-state (EOS) parameter ($w_{\rm DE}$) to cross the phantom divide
during the evolution of the Universe, an intriguing possibility
for which there is observational evidence from a variety of
sources (see \cite{Cai:2009zp} for a review).

The aforementioned models offer a satisfactory description of the
behaviour of dark energy and its observable features. However, the
dynamical nature of dark energy in these scenarios leads to the
``coincidence'' problem, which many authors have sought to resolve
by considering a coupling between dark energy and the other
components of the Universe. Thus, various forms of interacting
dark energy models have been constructed, including coupled
quintessence \cite{interacting} and interacting phantom models
\cite{Guo:2004vg}. Similarly, the dynamical nature of dark energy
makes it difficult to fulfil the observational requirement,
$w_{\rm DE}\approx-1$ \cite{obsdifficulties}, which has led to the
introduction of additional scalar fields. Inspired by similar
multi-field inflation \cite{multiinfaltion} and assisted inflation
constructions \cite{assistedinfaltion}, various forms of
assisted dark energy models have been considered
\cite{Barreiro:1999zs,assisted}.

One generalised class of the multi-field quintessence scenario
allows for a mixing of the fields' kinetic terms
\cite{vandeBruck:2009gp} (see also \cite{mixed}). Mixed kinetic
terms appear also in the non-trivial scalar field models based on
string theory constructions, studied in the framework of
generalised multi-field inflation \cite{Langlois:2008mn}.

In the present work we aim to combine the advantages of the quintom
scenario with those of assisted quintessence with mixed kinetic
terms. Thus, we construct a quintom model in which the kinetic
terms of the canonical and the phantom fields are mixed. Indeed,
as we see, the model exhibits features characteristic of both quintessence and phantom
models.

The plan of the work is as follows. In Sec. \ref{model} we
construct the quintom cosmological scenario with mixed kinetic
terms and present the formalism for its transformation into an
autonomous dynamical system. In Sec. \ref{phaseanalysis} we
perform the phase-space stability analysis and in Sec.
\ref{cosmimpl} we discuss the cosmological implications of the
results. Finally, our conclusions are presented in Sec.
 \ref{conclusions}.

\section{Quintom scenario with mixed kinetic terms}
\label{model}

Let us construct the quintom cosmological scenario with mixed
kinetic terms. Throughout the work we consider a flat
Robertson-Walker metric:
\begin{equation}\label{metric}
ds^{2}=dt^{2}-a^{2}(t){\bf{dx}}^2,
\end{equation}
with $a$ the scale factor.

We consider a model consisting of a canonical scalar field $\phi$,
a phantom field $\sigma$ and a mixed kinetic term proportional to
the parameter $\alpha$. Thus, the action, in units where $8\pi
G=1$ reads:
\begin{eqnarray}
S=\int d^4x \sqrt{-g}\left[\frac{R}{2} -
\frac{1}{2}g^{\mu\nu}\p_\mu\phi\p_\nu\phi +
 \frac{1}{2}g^{\mu\nu}\p_\mu\sigma\p_\nu\sigma -\right. \nonumber\\
\left. -\frac{\alpha}{2} g^{\mu\nu}\p_\mu\phi\p_\nu\sigma
 +V_\phi (\phi)+V_\sigma (\sigma) \right]+S_M,\
\end{eqnarray}
with $V_\phi (\phi)$ and $V_\sigma (\sigma)$ the corresponding
field potentials, and  $S_M$ the action for the matter (dark plus
baryonic) component of the Universe. We assume that the matter is
described by a perfect fluid with energy density $\rho_M$,
pressure $p_M$ and barotropic index $\gamma$. Thus,
$p_M=(\gamma-1)\rho_M$ and the matter equation-of-state parameter
is $w_M=\gamma-1$. Varying the action with respect to the metric
yields the Friedmann equations
\begin{eqnarray}
&&H^2 =
\frac{1}{3}\Big[\frac{1}{2}\dot\phi^2-\frac{1}{2}\dot\sigma^2+\frac{\alpha}{2}\dot\phi\dot\sigma+\ \ \ \ \ \ \ \ \ \ \ \ \ \ \ \ \ \ \ \ \ \ \ \ \nonumber\\
&&\ \ \ \ \ \ \ \ \ \ \ \ \ \ \ \ \  \ \ \ \ \ \ \ \ \ \ +V_\phi(\phi)+V_\sigma(\sigma)+\rho_M\Big], \label{FR1} \\
&&\dot H +\frac{3}{2}H^2=-\frac{1}{2}\Big[
\frac{1}{2}\dot\phi^2-\frac{1}{2}\dot\sigma^2+\frac{\alpha}{2}\dot\phi\dot\sigma-\ \ \ \ \ \ \ \ \ \ \ \ \ \ \ \ \ \ \nonumber\\
&&\ \ \ \ \ \ \ \ \ \ \ \ \ \ \ \ \ \ \ \ \ \ \ \ \ \ \ \ \
-V_\phi(\phi)-V_\sigma(\sigma)+p_M\Big],\label{FR2}
\end{eqnarray}
while the evolution equations for the two fields read
\begin{eqnarray}
\ddot\phi+3H\dot\phi+ \left(\frac{\p V_\phi}{\p\phi} +
\frac{\alpha}{2}\frac{\p V_\sigma}{\p\sigma}\right)
\left(1+\frac{\alpha^2}{4}\right)^{-1}= 0, \label{eq:phi_bac} \\
\ddot\sigma+3H\dot\sigma+ \left(-\frac{\p V_\sigma}{\p\sigma}
+\frac{\alpha}{2}\frac{\p
V_\phi}{\p\phi}\right)\left(1+\frac{\alpha^2}{4}\right)^{-1}= 0.
\label{eq:sigma_bac}
\end{eqnarray}

In quintom scenarios the dark energy is attributed to the
combination of the two scalar fields. In the case at hand, from
the Friedmann equations (\ref{FR1}),(\ref{FR2}) we
straightforwardly read the dark energy  density and pressure as
\begin{eqnarray}
\label{rhoDE}
 \rho_{DE}&=& \frac{1}{2}\dot\phi^2-\frac{1}{2}\dot\sigma^2+\frac{\alpha}{2}\dot\phi\dot\sigma +V_\phi(\phi)+V_\sigma(\sigma)\\
 p_{DE}&=& \frac{1}{2}\dot\phi^2-\frac{1}{2}\dot\sigma^2+\frac{\alpha}{2}\dot\phi\dot\sigma -V_\phi(\phi)-V_\sigma(\sigma),
 \label{pDE}
\end{eqnarray}
and thus for the dark energy equation-of-state parameter we
obtain
\begin{equation}
\label{wDE}
 w_{\rm DE}\equiv\frac{p_{DE}}{
\rho_{DE}}=\frac{\frac{1}{2}\dot\phi^2-\frac{1}{2}\dot\sigma^2+\frac{\alpha}{2}\dot\phi\dot\sigma
-V_\phi(\phi)-V_\sigma(\sigma)}{
\frac{1}{2}\dot\phi^2-\frac{1}{2}\dot\sigma^2+\frac{\alpha}{2}\dot\phi\dot\sigma
+V_\phi(\phi)+V_\sigma(\sigma)}.
\end{equation}
Finally, using (\ref{rhoDE}),(\ref{pDE}) and
(\ref{eq:phi_bac}),(\ref{eq:sigma_bac}),  we can easily acquire
the dark energy evolution equation in fluid terms, which takes the
expected form
\begin{equation}
\label{rhot}
 \dot{\rho}_{DE}+3 H(\rho_{DE}+p_{DE})=0.
\end{equation}
We mention that in the case $\alpha=0$ the above scenario
coincides with the usual quintom one \cite{quintom}.

\section{Phase-space analysis}
\label{phaseanalysis}

In order to perform the phase-space and stability analysis of the
quintom model at hand, we have to transform the aforementioned
dynamical system into its autonomous form
\cite{expon,Copeland:1997et,DEreview}. This can be achieved by introducing
the auxiliary variables:
\begin{eqnarray}
   x_\phi=\frac{\dot{\phi}}{\sqrt{6}H}&,&\ \  x_\sigma=\frac{\dot{\sigma}}{\sqrt{6}H},\nonumber\\
 y_\phi=\frac{\sqrt{V_\phi(\phi)}}{\sqrt{3}H}&,&\ \  y_\sigma=\frac{\sqrt{V_\sigma(\sigma)}}{\sqrt{3}H}, \label{auxilliary}
\end{eqnarray}
together with $N=\ln a$. It is easy to see that for every
quantity {\it F} we acquire $\dot{F}=H\frac{dF}{dN}$.
 Using these
variables, from (\ref{rhoDE})  we can calculate the dark energy
density parameter,
\begin{equation}
 \Om\equiv\frac{\rho_{DE}}{3H^{2}}=x_\phi^2-x_\sigma^2+\alpha
 x_\phi x_\sigma+y_\phi^2+y_\sigma^2
 \label{OmegaDE},
\end{equation}
and from (\ref{wDE}) the dark energy equation-of-state parameter,
\begin{equation}
\label{wDE2} w_{\rm DE}=\frac{ x_\phi^2-x_\sigma^2+\alpha
 x_\phi x_\sigma-y_\phi^2-y_\sigma^2}{ x_\phi^2-x_\sigma^2+\alpha
 x_\phi x_\sigma+y_\phi^2+y_\sigma^2}.
\end{equation}

The final step is the  consideration of specific potential forms.
Following works on assisted quintessence \cite{assisted} and
assisted inflation \cite{assistedinfaltion} scenarios, together
with various other  cosmological models
\cite{expon,Copeland:1997et,Setare:2008si}, we consider
exponential potentials of the form,
\begin{equation}
V_\phi(\phi)= e^{-\lambda\phi},\hspace{1cm}
V_\sigma(\sigma) = e^{-\mu\sigma}.
\end{equation}
 Note
that equivalently, but more generally, we could consider
potentials satisfying $\lambda=-\frac{1}{\kappa
V(\phi)}\frac{\partial V(\phi)}{\partial\phi}\approx {\rm const}$,
$\mu=-\frac{1}{\kappa V(\sigma)}\frac{\partial
V(\sigma)}{\partial\sigma}\approx {\rm const}$ (for example this
relation is valid for arbitrary but nearly flat potentials
\cite{Scherrer:2007pu}). Finally, without loss of generality, in
this work we consider $\lambda$ and $\mu$ to be positive, since
the negative case  alters only the direction towards which the fields
roll, and can always be transformed into the  positive case by
changing the signs of the fields themselves.

Let us now perform the phase-space analysis of the
model. In general, having transformed the cosmological system into
its autonomous form
 $\textbf{X}'=\textbf{f(X)}, $ where $\textbf{X}$ is the column
vector constituted by the auxiliary variables, \textbf{f(X)} the
corresponding  column vector of the autonomous equations, and
prime denotes a derivative with respect to $N=\ln a$, the critical
points $\bf{X_c}$ are extracted satisfying $\bf{X}'=0$. In order
to determine the stability properties of these critical points we
expand around $\bf{X_c}$, setting $\bf{X}=\bf{X_c}+\bf{U}$ with
$\textbf{U}$ the perturbations of the variables considered as a
column vector. Up to first order we acquire $
\textbf{U}'={\bf{Q}}\cdot \textbf{U}, $ where the matrix ${\bf
{Q}}$ contains the coefficients of the perturbation equations.
Thus, for each critical point, the eigenvalues of ${\bf {Q}}$
determine its type and stability.

Following this method and using the auxiliary variables
(\ref{auxilliary}), the cosmological equations of motion
(\ref{FR1}), (\ref{FR2}), (\ref{eq:phi_bac}) and
(\ref{eq:sigma_bac}) become
\begin{eqnarray}
x_\phi'
&=&-3x_\phi+\sqrt{\frac{3}{2}}\left(1+\frac{\alpha^2}{4}\right)^{-1}
\left(  \lambda y_\phi^2  + \frac{\alpha}{2}\mu y_\sigma^2 \right)
+ x_\phi T, \nonumber
\\
x_\sigma'
&=&-3x_\sigma-\sqrt{\frac{3}{2}}\left(1+\frac{\alpha^2}{4}\right)^{-1}
\left(  \mu y_\sigma^2 - \frac{\alpha}{2}\lambda y_\phi^2 \right)+
x_\sigma T, \nonumber
 \\
y_\phi'&=&-\sqrt{\frac{3}{2}}\lambda x_\phi y_\phi+y_\phi T,
\nonumber\\
y_\sigma'&=&-\sqrt{\frac{3}{2}}\mu x_\sigma y_\sigma+y_\sigma T,
\label{eq:Xia}
\end{eqnarray}
where
\begin{eqnarray}
 T&=&\frac{3}{2}\Big[  2 x_\phi^2 -2 x_\sigma^2+2\alpha x_\phi
x_\sigma
 \nonumber\\
&{}& +\gamma (1-x_\phi^2+x_\sigma^2-\alpha x_\phi
x_\sigma-y_\phi^2-y_\sigma^2)\Big ].
\end{eqnarray}

The critical points $(x_{\phi c},x_{\sigma c}, y_{\phi c},
y_{\sigma c})$ of the autonomous system (\ref{eq:Xia}) are
obtained by setting the left hand sides of the equations to zero.
The real and physically meaningful (i.e. corresponding to
$y_\phi,y_\sigma>0$, and $0\leq\Omega_{DE}\leq1$) points are
presented in Table \ref{table:points}.


\begin{table*}[tb!]

\footnotesize

\begin{tabular*}{1\textwidth}{@{\extracolsep{\fill}} | c || cc | cc | cc | cc |}
\hline
&  $x_{\phi c}$ & & $x_{\sigma c}$ & & $y_{\phi c}$&  & $y_{\sigma c}$ &  \\
\hline & & & & & & & & \\ [-1em] \hline
A  & 0 & &  0 & &  0 & &  0 & \\

\hline
& & & & & & & & \\ [-1em]
B & $-\tfrac{1}{2}x_\sigma \alpha \pm \tfrac{1}{2}\sqrt{x_\sigma^2(\alpha^2+4)+4}  $ & & $x_\sigma$ & & 0 & & 0 & \\

\hline
& & & & & & & & \\ [-1em]
D1 & $ \frac{\sqrt{6}\mu \alpha}{3(4+\alpha^2)} $ & &
 $  -\frac{2\sqrt{6}\mu}{3(4+\alpha^2)} $ & &
 0 & &
$\frac{\sqrt{6}}{3(4+\alpha^2)}\sqrt{(4+\alpha^2)(6+\tfrac{3}{2}\alpha^2+\mu^2)} $ & \\

\hline
& & & & & & & &  \\ [-1em]
C2  &
 $\frac{\gamma \sqrt{6}}{2\lambda}$ & &
 $\frac{\alpha\gamma \sqrt{6}}{4\lambda}$ & &
$\frac{\gamma \sqrt{6}}{4\lambda}\sqrt{(4+\alpha^2)(\tfrac{2}{\gamma}-1)}$ & &
0 &\\

\hline
 & & & & & &  & & \\ [-1em]
D2  & $  \frac{2\sqrt{6}\lambda}{3(4+\alpha^2)} $ & &
$ \frac{\sqrt{6}\lambda \alpha}{3(4+\alpha^2)} $ & &
$\frac{\sqrt{6}}{3(4+\alpha^2)}\sqrt{(4+\alpha^2)(6+\tfrac{3}{2}\alpha^2-\lambda^2)} $ & &
0 & \\

\hline
& & & & & & & &  \\ [-1em]
 E & $ \sqrt{\frac{3}{2}}\frac{\gamma}{\lambda} $ & &
 $\sqrt{\frac{3}{2}}\frac{\gamma}{\mu} $ & &
$\sqrt{\frac{3}{2}}\frac{\gamma}{\mu\lambda}\sqrt{\mu(\mu+\frac{\alpha}{2}\lambda)(\frac{2}{\gamma}-1)}$ & &
$\sqrt{\frac{3}{2}}\frac{\gamma}{\mu\lambda}\sqrt{\lambda(\frac{\alpha}{2}\mu-\lambda)(\frac{2}{\gamma}-1)} $ &  \\
\hline
& & & & & & & &  \\ [-1em]
F & $ \frac{\Pi}{ \sqrt{6}\lambda} $  & &
$ \frac{\Pi}{ \sqrt{6}\mu} $   & &
$ \frac{1}{\mu\lambda} \sqrt{\Pi \mu \left(\mu+\frac{\alpha}{2}\lambda\right)\left(1-\frac{\Pi}{6}\right) }$  & &
$ \frac{1}{\mu\lambda} \sqrt{\Pi \lambda \left(\frac{\alpha}{2}\mu -\lambda \right)\left(1-\frac{\Pi}{6}\right) }$ &  \\

\hline
\end{tabular*}
\caption[crit]{\label{table:points} The real and physically
meaningful critical points of the autonomous  system
(\ref{eq:Xia}).
The parameter $\Pi$ is defined in (\ref{eq:lameff}).
}
\end{table*}

The $4\times4$ matrix ${\bf {Q}}$ of the linearised perturbation
equations of the system (\ref{eq:Xia}) is shown in the Appendix.
Thus, for each critical point of Table \ref{table:points} we
examine the sign of the real part of the eigenvalues of ${\bf
{Q}}$ to determine the type and stability of the point. In Table
\ref{table:stability} we present the results of the stability
analysis. In addition, for each critical point we calculate the
values of $\Omega_{DE}$ [given by  (\ref{OmegaDE})] and of
 $w_{\rm DE}$ [given by (\ref{wDE2})].
Finally, we mention that the labelling of the critical points
follows the convention  used
 in  \cite{vandeBruck:2009gp}. That is, since in a two-field
 system
 there are critical points in which the potential term of one of the fields is
 zero,
  the connection between these points
 is made more explicit. For instance,
  in the points {\it D1} and {\it D2} (for which $\Om=1$)
 only one of the fields' potential
 terms contributes to the energy density.
 Note that in this model (unlike that in \cite{vandeBruck:2009gp}) the solution
 satisfying $\bf{X}'=0$ with $y_{\phi}=0$ corresponding to {\it C1} is unphysical.

The stability of the critical points can be summarised as follows:

\begin{itemize}

\item {\bf Point {\textit A}}

This is a trivial solution, corresponding to the fluid  dominated
point {\it F} in \cite{Guo:2004fq}, where the kinetic and potential
components of both the phantom and the quintessence fields are
negligible. It exists for all values of $\alpha$, $\lambda$ and
$\mu$. The eigenvalues are \eqa
&e_1&=e_2=\tfrac{3}{2}\gamma>0, \nonumber \\
& e_3&=e_4= -\tfrac{3}{2}\left(2-\gamma  \right)<0, \eqae
so this is a saddle point.

\item {\bf Point {\textit B}}

The two points indicated by the  $\pm$ are similar to point B in
\cite{vandeBruck:2009gp}. When $\alpha=0$ (point {\it K} in
\cite{Guo:2004fq}), there is a continuous locus of points that
describes a hyperbola in the $(x_\phi,x_\sigma)$ plane. When
$\alpha\ne 0$ the locus still describes a hyperbola, but the
asymptotes are no longer orthogonal. The eigenvalues are
\eqa
e_1 &=& 3(2-\gamma)>0,  \nn \\
e_2 &=& 0, \nn \\
e_3 &=& 3-  \tfrac{5\sqrt{6}}{4}\mu x_\sigma \nn\\
&+& \tfrac{3\sqrt{6}}{4}\lambda\left[ -\tfrac{1}{2}x_\sigma \alpha \pm \tfrac{1}{2}\sqrt{x_\sigma^2(\alpha^2+4)+4}  \right],  \nn \\
e_4 &=& 3+\tfrac{3\sqrt{6}}{4}\mu x_\sigma \nn\\
&-&\tfrac{5\sqrt{6}}{4}\lambda\left[ -\tfrac{1}{2}x_\sigma \alpha
\pm \tfrac{1}{2}\sqrt{x_\sigma^2(\alpha^2+4)+4}  \right], 
\eqae
where the $\pm$ in  $e_3$ and $e_4$ corresponds to the particular
solution considered.  $e_1$ is positive, therefore the point is
unstable.

\item {\bf Point {\textit D1}}

This takes a similar form to point {\it D1}  in
\cite{vandeBruck:2009gp}. (The $y_\sigma$ value in the solution
corresponding to {\it C1} in \cite{vandeBruck:2009gp} has a factor of
$\sqrt{(4+\alpha^2)(1-2/\gamma)}$ so there are no real values of
$\alpha$ for which the point exists.) In the standard quintom case
($\alpha=0$), {\it D1} is a phantom-dominated point (point {\it P} in the
notation of \cite{Guo:2004fq}). In that case, as in the present
one, the term under the square root in $y_{\sigma c}$ is positive
and $\Om=1$, so the point exists for all values of the parameters.
However, in the case at hand, the eigenvalues read
 \eqa
e_{1,2} &=&  -\frac{2\mu^2}{ (4+\alpha^2) }-3 \le 0, \nn \\
e_3 &=&  -\frac{ 4\mu^2  }{ (4+\alpha^2) } - 3\gamma \le 0, \nn \\
e_4 &=&  -\frac{ \mu (2\mu + \alpha \lambda) }{ (4+\alpha^2) } ,
\eqae 
so the point is stable only if $(2\mu + \alpha \lambda) >0$,
which is trivially satisfied when $\alpha\ge 0 $.


\renewcommand{\tabularxcolumn}[1]{m{#1}}
\begin{table*}[tb!]

\centering
\newcolumntype{R}{>{\center}X}
\begin{tabularx}{\textwidth}[b]{ | c || R | R | c | c |  }
\hline
& Existence & Conditions for Stability & $\Om$ & $w_{\rm DE}$ \\
\hline
 & & &  & \\ [-1em]
\hline
  & & &  & \\ [-1.7em]
A & all $\mu$,$\lambda$,$\alpha$ & unstable & 0 & undefined \\
\hline
  & & &  & \\ [-1.7em]
B & all $\mu$,$\lambda$,$\alpha$ & unstable & 1 & 1 \\

\hline
 & & &  & \\ [-1.4em]
D1 &
all $\mu$,$\lambda$,$\alpha$ &
$\mu> - \alpha\lambda/2$ &
 1 &
 $-1-\frac{4\mu^2}{3(4+\alpha^2)}$ \\

 \hline
  & & &  & \\ [-1.7em]
C2 & $ \lambda \ge \sqrt{\frac{3\gamma}{4}(4+\alpha^2)}$ &
$\lambda < \alpha\mu/2$ &
$ \frac{3\gamma}{4\lambda^2}(4+\alpha^2) $ &
$w_M$  \\
\hline
 & & &  & \\ [-1.7em]
D2 &
 $ \lambda \le \sqrt{\frac{3}{2}(4+\alpha^2)}$ &
$\lambda < min\left[ \alpha\mu/2,\sqrt{\frac{3\gamma}{4}(4+\alpha^2)} \right] $&
 1 &
 $-1+\frac{4\lambda^2}{3(4+\alpha^2)}$ \\

  \hline
  & & &  & \\ [-1.7em]
E & $\alpha\mu-2\lambda >0\;\;$; $2\mu+\alpha\lambda >0\;\;$; $ \Pi \ge 3\gamma $ & unstable &
 $3\gamma/ \Pi$ &
$w_M$  \\

  \hline
    & & &  & \\ [-1.7em]
 & $\alpha\mu-2\lambda >0\;\;$; $ 0<\Pi<6$  &  unstable  & &  \\

F & & &  1 & $ -1+\Pi/3 $ \\
    & & &  & \\ [-1.7em]
 & $2\mu+\alpha\lambda <0$ &  stable & & \\

\hline
\end{tabularx}
\caption{The properties of the critical points of the autonomous
system (\ref{eq:Xia}). The parameter $\Pi$ is defined in (\ref{eq:lameff}).
Note that there are two disconnected regions in the parameter
space for which point {\it F} is physical.}
  \label{table:stability}
\end{table*}


\item {\bf Point {\textit C2}}

The points {\it C2} and {\it D2} are similar to {\it D1}, in that only one of the
fields' potential energy contributes to the total energy density;
in {\it D1} this was the phantom field, while here it is the
quintessence one. This means that there is both a scaling solution
({\it C2}) and a field-dominated solution ({\it D2}), for which $\Om=1$. The
density parameter for {\it C2} is $\Om=
\frac{3\gamma}{4\lambda^2}(4+\alpha^2) $, so the condition $\Om\le
1$ means that $ \lambda \ge \sqrt{\frac{3\gamma}{4}(4+\alpha^2)}$
must be satisfied if the point is to be physical. The eigenvalues
are
\eqa
&e_1&= -\tfrac{3}{2}(2-\gamma)<0, \nn \\
&e_2 &= \tfrac{3\gamma}{4\lambda}(2\lambda-\alpha\mu), \nn \\
&e_{3,4}&= -\tfrac{3(2-\gamma)}{4}  \nn\\
&\times& \left[ 1 \pm \sqrt{ \frac{
6\gamma^2(4+\alpha^2)+\lambda^2(2-9\gamma)}{\lambda^2(2-\gamma)} }
\right].
 \eqae
  $e_3$ and $e_4$ are always negative as the
condition for existence gives
\[
 6\gamma^2(4+\alpha^2)+\lambda^2(2-9\gamma) \le  \lambda^2(2-\gamma).
\]
Thus, the point is stable if $(2\lambda-\alpha\mu)<0$,  so unlike
the quintom case (cf. point T in \cite{Guo:2004fq}), with a
sufficiently large value of $\alpha\mu$ one obtains a stable
scaling solution in this system.

\item {\bf Point {\textit D2}}

This field-dominated solution is similar to {\it D1} in
\cite{vandeBruck:2009gp} and reduces to point {\it S} in
\cite{Guo:2004fq} when $\alpha=0$. The existence condition, $
\lambda \le \sqrt{\frac{3}{2}(4+\alpha^2)}$, is derived from the
requirement that $y_{\phi c}\in\mathbb{R}$. The eigenvalues are
\begin{equation*}
e_1=  \frac{ \lambda (2\lambda - \alpha \mu) }{ (4+\alpha^2) }      \hspace{0.4cm}     e_2 =  \frac{ 4\lambda^2  }{ (4+\alpha^2) } - 3\gamma \nn
\end{equation*}
\eq e_{3,4}=  \frac{2\lambda^2}{ (4+\alpha^2) }-3 \le 0, \eqe
where the existence condition ensures that $e_{3,4}$ are negative.
If the point is to be stable both $e_1$ and $e_2$ must be
negative, which is satisfied if
\[
\lambda< min\left[ \alpha\mu/2, \sqrt{\tfrac{3\gamma}{4}{(4+\alpha^2)}} \right].
\]
Once again, one can choose values of $\mu$ and $\alpha$ so that
this point is stable, which is not the case in the standard
quintom model.

\item {\bf Point {\textit E}}

The points {\it E} and {\it F} are characterised by the fact that the
potential terms  for both the phantom and the quintessence field
are non-zero. We mention that when $\alpha=0$  these points do not
exist, since it is not possible to choose values for $\lambda$ and
$\mu$ such that $y_{\phi c}, y_{\sigma c}\in\mathbb{R}$. Point {\it E}
is a scaling solution, with density parameter
\[
\Om=\frac{3\gamma}{\lambda^2\mu^2}\left( \mu^2-\lambda^2+a\mu\lambda \right).
\]
Following \cite{vandeBruck:2009gp}, we can rewrite this quantity
in terms of an effective exponent $\Pi$ (which in the present case
can be negative too, contrary to \cite{vandeBruck:2009gp}) defined
by
 \eq \label{eq:lameff} \frac{1}{\Pi} =
\frac{1}{\lambda^2}-\frac{1}{\mu^2}+\frac{\alpha}{\mu\lambda},
\eqe
 which leads to  the existence condition  $ \Pi \ge 3\gamma $.
Note that, as $\mu,\lambda \ge 0$, \eq (\alpha\mu-2\lambda)>0
\Rightarrow (2\mu+\alpha\lambda)>0.
\label{eq:uneq1}
 \eqe
  Thus, to
ensure that $y_{\phi c}$ and $y_{\sigma c}$ are real, one must
additionally impose $(\alpha\mu-2\lambda)>0$. The eigenvalues
are
\eqa
e_{1,2}&=&-\tfrac{3}{4}(2-\gamma) \nn\\
&\times& \left[ 1\pm \sqrt{ \frac{24\gamma^2/\Pi+2-9\gamma }{(2-\gamma)} }\right], \label{eq:Eeigen12} \\
e_{3,4}&=&-\tfrac{3}{4}(2-\gamma)\nn\\
&\times& \left[ 1\pm\sqrt{
1-\frac{8\gamma(2\mu+\alpha\lambda)(2\lambda-\alpha\mu)}{\lambda\mu(2-\gamma)(4+\alpha^2)}
}  \right].  \label{eq:Eeigen34}
 \eqae
  In terms of $\Pi$  the
existence condition reads $  \frac{24\gamma^2}{\Pi}+2-9\gamma \le
(2-\gamma)$; substituting this into (\ref{eq:Eeigen12}) leads to
$e_{1,2} \le -\tfrac{3}{4}(2-\gamma)(1\pm 1) \le 0$. However,
looking at   the fraction under the square root in
(\ref{eq:Eeigen34}), we observe that
\[
1-\frac{8\gamma(2\mu+\alpha\lambda)(2\lambda-\alpha\mu)}{\lambda\mu(2-\gamma)(4+\alpha^2)}>1.
\]
Thus, either $e_3$ or $e_4$ must be positive, and therefore the
critical point is unstable.

\item {\bf Point {\textit F}}

In this (field-dominated) point we observe the presence of
square-root terms in $y_{\phi c}$ and $y_{\sigma c}$ . Thus, for
these values to be real we require
 \eqa
 \Pi\left(1-\frac{\Pi}{6}\right) \left(2\mu+\alpha\lambda\right)&>&0, \nn \\
\Pi \left(1-\frac{\Pi}{6}\right) \left(\alpha\mu -2\lambda \right)&>&0. \nn
 \eqae
Using (\ref{eq:lameff}) and (\ref{eq:uneq1}), and noting that \eq
(2\mu+\alpha\lambda)<0  \Rightarrow (\alpha\mu-2\lambda)<0
\label{eq:uneq2}, \eqe and \eqa
(2\mu+\alpha\lambda)<0  &\Rightarrow& \mu^2-\lambda^2+a\mu\lambda < 0 \nn \\
 &\Rightarrow& \Pi <0, \label{eq:uneq3}
\eqae we conclude that there exist two possible cases: \eqa
 &(a)&  \alpha\mu-2\lambda >0; \hspace{1em} 0<\Pi<6,
 \label{eq:case(a)}\\
 &(b)&  2\mu+\alpha\lambda <0.
 \label{eq:case(b)}
\eqae
 The eigenvalues read
 \eqa
 e_1&=& -3\left(1-\frac{\Pi}{6} \right), \nn \\
 e_2 &=& \Pi-3\gamma, \nn\\
e_{3,4} &=& -\frac{3}{2} \left(1-\frac{\Pi}{6}\right) \nn\\
 &\times&\left[ 1\pm \sqrt{ 1-\frac{4\Pi}{3\lambda\mu}\frac{(2\mu+\alpha\lambda)(2\lambda-\alpha\mu)}{(4+\alpha^2)\left(1-\tfrac{\Pi}{6}   \right)} }  \right].
\eqae
\begin{figure}[th!]
\centering \subfigure[] {
\includegraphics[width=8.2cm]{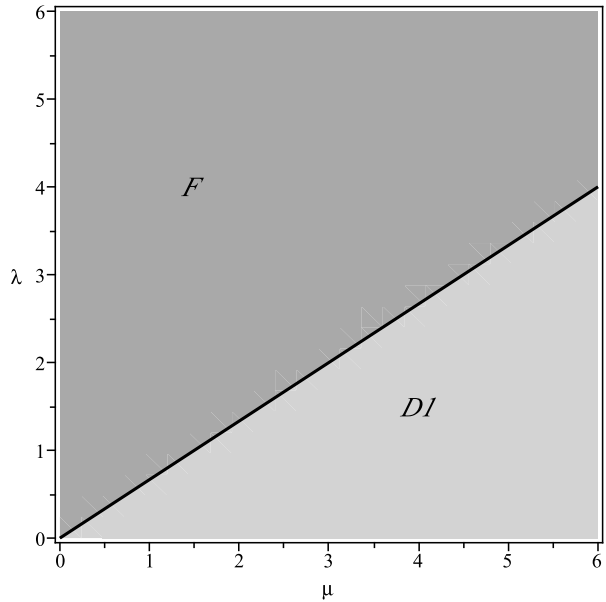}
} \subfigure[] {
\includegraphics[width=8.2cm]{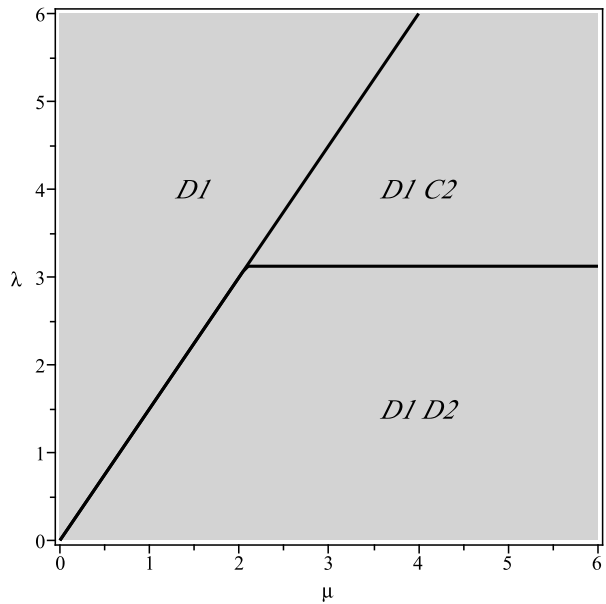}
} \caption{The parameter space showing the stability of the
critical points for  (a) $\alpha=-3$ and (b) $\alpha=3$, with
$\gamma=1$. The light shaded area indicates the region where {\it D1} is
stable and the dark shaded area indicates the region where {\it F} is
stable. There is a stable point with $w_{\rm DE}<-1$ at each region in
the parameter space. The area marked in graph (b) indicates the
region where {\it C2} and {\it D2} are stable.} \label{fig:stable_space}
\end{figure}
  In   case (a), $\left(1-\frac{\Pi}{6} \right)>0$, thus $e_1<0$.
However, in a similar way to the analysis of point {\it E}, the term
under the square root in the expression for $e_{3,4}$ is greater
than 1 so either $e_3$ or $e_4$ is positive and the point is
unstable. In case (b), the term under the square root is less than
1 so $e_{3,4}<0$.  $e_2<0$ is trivially satisfied [using
(\ref{eq:uneq3})] and, again, $\left(1-\frac{\Pi}{6} \right)>0$,
therefore $e_1<0$. Thus, the point is stable whenever the
condition (\ref{eq:case(b)}) is satisfied. Finally, we mention
that this separation of the parameter space is a novel
feature of the scenario at hand, which is not present in the
corresponding case of two canonical fields
\cite{vandeBruck:2009gp}.
\end{itemize}

The stability regions for the points  are plotted in Fig.
\ref{fig:stable_space}. In the standard quintom case ($\alpha=0$),
there is only one stable critical point that exists for all values
of $\mu$ and $\lambda$, corresponding to the phantom point {\it D1}.
In
the case $\alpha <0$  the parameter space is fractured, thus,
although there is a stable point with $w_{\rm DE}<-1$ at each
parameter-space region, the precise value of $w_{\rm DE}$ for
particular values of $\mu$ and $\lambda$ depends on whether point
{\it D1} or {\it F} is stable.
In the case where $\alpha>0$, the phantom
dominated point {\it D1} is always stable. It is interesting, however,
that there are also regions where {\it C2} (a scaling solution) and {\it D2}
(a quintessence-dominated solution), each with $w_{\rm DE}>-1$, are
stable.
In these regions the initial conditions will determine the
cosmological evolution.
\begin{figure}[t]
\centering
\includegraphics[width=8.2cm]{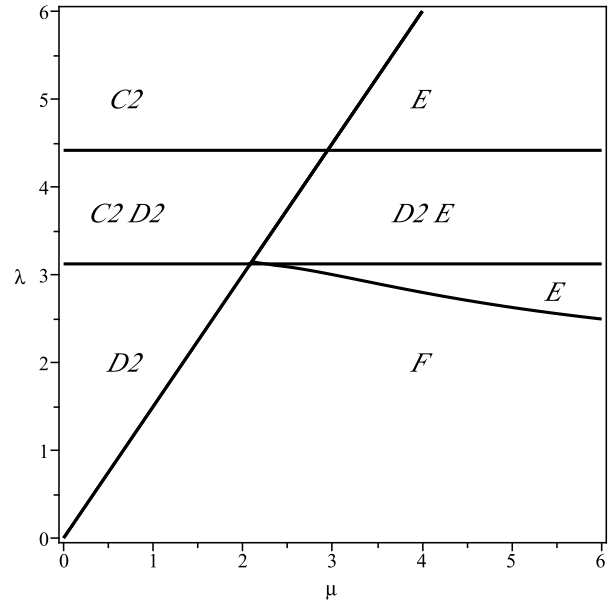}
\caption{Regions in the parameter space where the critical points
exist but are unstable for $\alpha=3$ (with $\gamma=1$).}
\label{fig:unstable_space}
\end{figure}

As we have seen, point {\it F}  exists in two disconnected regions in
the parameter space: a stable region where $\alpha<0$ and an
unstable region where $\alpha>0$. This can be understood as a
fracture of the unstable regions in the parameter space, depicted
in Fig. \ref{fig:unstable_space}. In the standard quintom case,
there are two overlapping regions corresponding to {\it C2} and {\it D2}. When
$\alpha\mu-2\lambda>0$  the points {\it E} and {\it F} are present and
encroach upon this region. As one might expect, in both cases,
when the exponent of the quintessence field is large, the scaling
solution does exist (although it is not always stable).

\section{Cosmological implications}
\label{cosmimpl}

Having performed the complete phase-space analysis of the model,
 we can discuss the corresponding cosmological behaviour. A
general remark is that the phenomenology is different from both
the standard ($\alpha=0$) quintom scenario
\cite{Guo:2004fq,quintom} and from the case of quintessence with
mixed kinetic terms \cite{vandeBruck:2009gp}. In particular, there
are late-time solutions describing both quintessence and phantom
dominated universes, with the crucial observable quantity being
the dark energy equation-of-state parameter $w_{\rm DE}$ (which is
greater than $-1$ in the former case and less than $-1$ in the
latter).

The points {\it A}, {\it B} and {\it E} are unstable, and thus cannot be
late-time solutions of the Universe. {\it A} corresponds to a
matter-dominated universe, while {\it B} represents a
dark energy-dominated  one, with the stiff value $w_{\rm DE}=1$. {\it E}
is a scaling solution. Finally, in one of the two disconnected
areas in which {\it F} is physical, when $\alpha\mu-2\lambda >0$
and $ 0<\Pi<6$, which corresponds to $w_{\rm DE}>-1$, the point is
always unstable.

As usual, the stable critical points can be classified as
field-dominated ($\Om=1$) or scaling solutions ($w_{\rm DE}=0$ in the
presence of matter). Since these features are incompatible with
observations, we deduce that the late-time behaviour of the
scenario considered here cannot provide a satisfactory description of the
Universe at the present time. However, the fact that dark energy
has only recently begun to dominate the energy density, after an
extended matter-dominated era, suggests that the Universe may have yet to
reach an attractor solution. In this case, one can
consider our Universe as evolving towards a field dominant
attractor, although the problem with initial conditions is still an
issue \cite{LopesFranca:2002ek}.

In order to treat the coincidence problem one must explain
why the present dark energy and matter density parameters are of
the same order of magnitude. Alternatively, one can ask why dark
energy has only recently become dynamically important, which is a
problem of initial conditions. This can be alleviated if
$\rho_{DE}$ scales with the dominant matter component, as the
discrepancy between the initial energy densities of the various
components of the Universe does not have to be so large.

In contrast to the simple quintom scenario with $\alpha=0$,
 there is a stable scaling solution
 in this model, viz. point {\it C2}, which is an extension of the unstable point
T in \cite{Guo:2004fq}. However, as the present model stands,
there is no mechanism to effect a transition between the scaling
regime and field-dominant solution without fine-tuning the initial
conditions.

Another novel feature of the extended quintom scenario is a stable
field-dominated late-time solution with $w_{\rm DE}>-1$ (point {\it D2})
which is characterised by the fact that the contribution of the
potential of the phantom field to the energy density is zero. Like
{\it C2}, this is an extension of an unstable point in the simple
quintom scenario. The deviation from $w_{\rm DE}=-1$ is
$4\lambda^2/[3(4+\alpha^2)]$ so one requires either the coupling
to be large or the quintessence potential to be sufficiently flat
for this solution to be compatible with observations.

Let us make an important comment concerning {\it C2} and {\it D2}. As
can be seen in Table \ref{table:points}, both
points have $x_{\sigma c}>0$ in the regions in which they are stable
 (we consider positive
potential exponents $\lambda,\mu>0$).
However, since phantom
fields generally do not roll down their potential slopes, one would normally expect that in a
stable late-time solution, the rate of change of the phantom field, that is
$x_{\sigma c}$, should be negative (in a stationary solution) or
zero (in a static solution). Indeed, for the simple quintom case
($\alpha=0$) we do obtain $x_{\sigma c}=0$. Interestingly
enough, the positive coupling of the kinetic terms of the two
fields makes the quintessence field drag the phantom field
down its potential (i.e. the opposite to ``normal'' phantom behaviour).
This possibility is a novel feature of the present scenario.

The point {\it D1} is the analogue of {\it D2} and is stable for the
larger part of the parameter space (it is always stable for
$\alpha\geq0$, while for $\alpha<0$ it can still be stable for
$\mu>-\alpha\lambda/2$). $w_{\rm DE}$ depends only on the values of $\mu$ and $\alpha$ and
approaches $-1$ if the potential is flat or the coupling is large,
in a similar manner to the quintessence-dominated point, {\it D2}.
Furthermore, it is interesting to note in this case there is a situation similar to
the one described in the previous paragraph.
In particular, for
$\alpha<0$ we observe that $x_{\phi c}<0$, so, contrary to expectations,
there is a stable late-time solution where the quintessence
field rolls up
its potential slope. Again, this is due to
the coupling of the kinetic terms: the phantom field feeds
the upward rolling of the quintessence field.

From the discussion of the above two paragraphs we deduce
that the kinetic coupling leads to new stable solutions.
Additionally, it could lead to a very interesting evolution towards
these late-time solutions. In particular, if the kinetic
terms are initially small, the effect of the coupling is downgraded,
and the fields can move away from these stable points. With
the increase of their kinetic energies, the coupling becomes more important
and they will start to move towards them.

When point {\it D1} is unstable, the critical point {\it F}, in one of the
two disconnected areas for which it is physical, viz.
$2\mu+\alpha\lambda <0$, is stable. It corresponds to a universe
completely dominated by dark energy, behaving like a phantom, with
the EOS determined by the parameter $\Pi$. In this case, the
quintessence and phantom fields combine to give a solution with
$w_{\rm DE}<-1$. This is the analogue of point {\it F} in
\cite{vandeBruck:2009gp} and, like the unstable point {\it E},
does not exist in the standard quintom case.

\begin{figure}[ht]
\centering \subfigure[] {
\includegraphics[width=8.6cm]{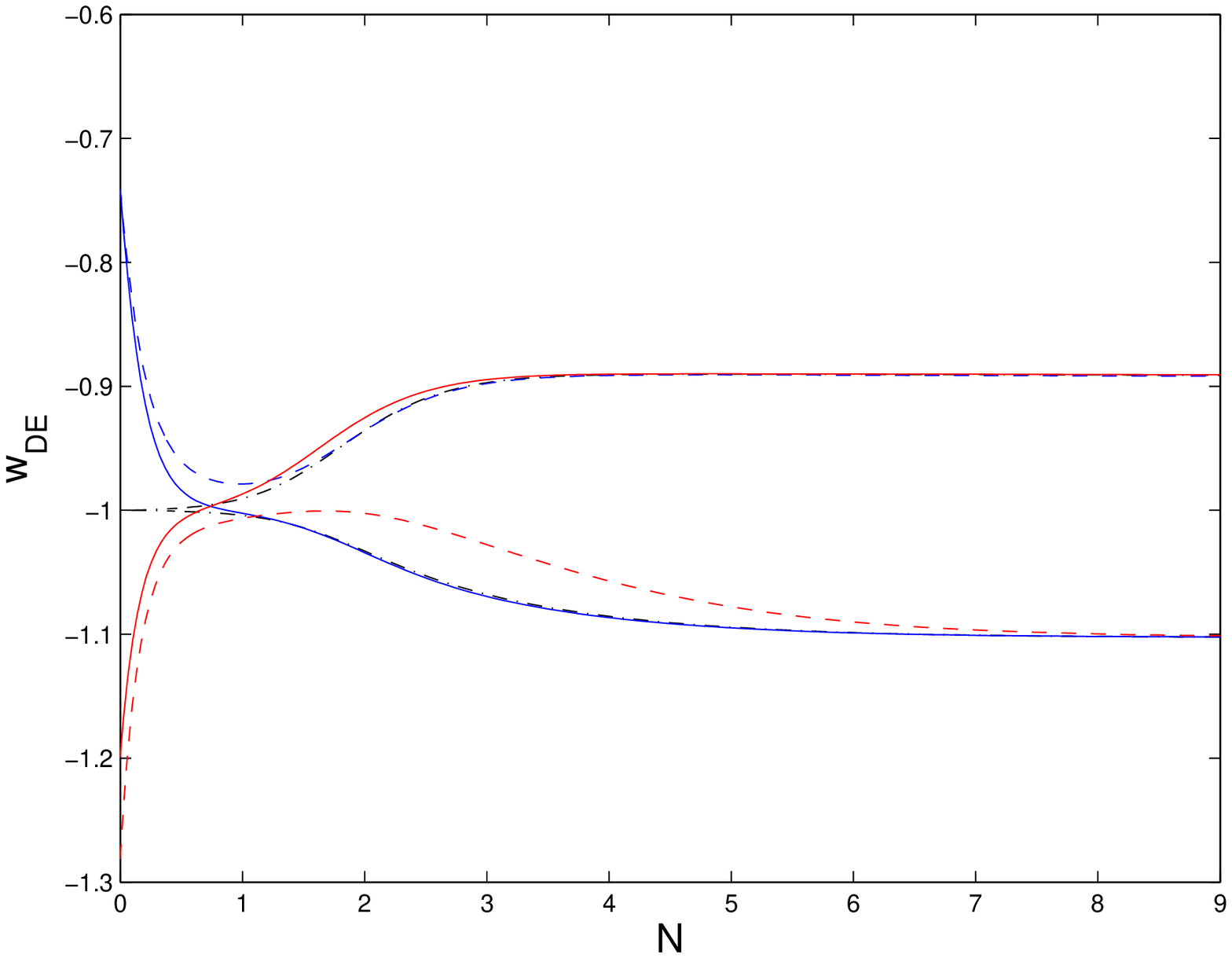}
} \subfigure[] {
\includegraphics[width=8.6cm]{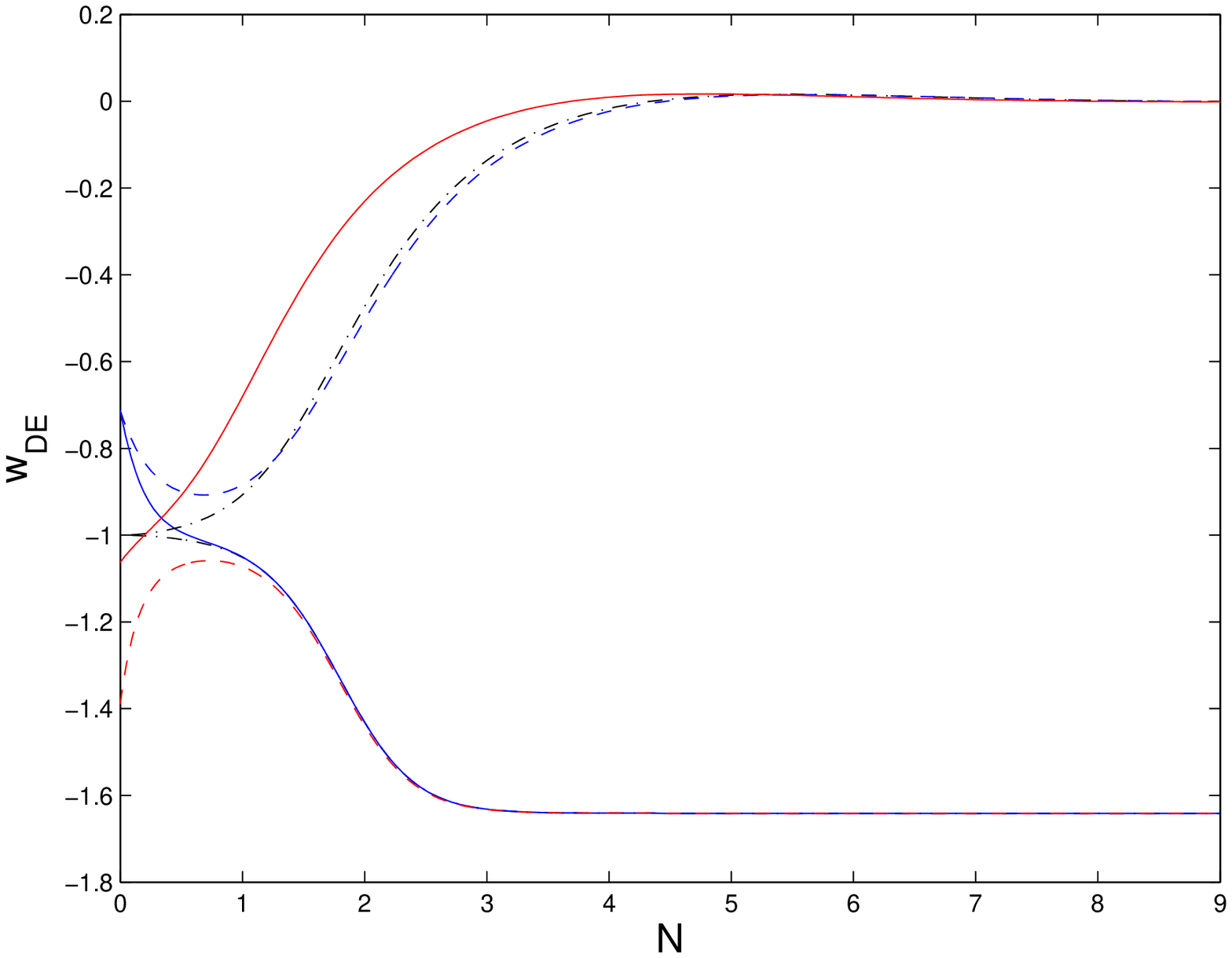}
} \caption{Numerical integration of the system with
(a) $\mu=1$, $\lambda=1$, $\alpha=3$ and (b) $\mu=2.5$,
$\lambda=3.5$, $\alpha=3$ with initial conditions chosen so that
the matter field (with $w_M=0$) initially dominates. A variety of
plots with different initial conditions are shown: the
colour-coding is for visual clarity. In case (a) points {\it D1} and
{\it D2} are stable; one can see that there are solutions that cross
the phantom divide to reach the late-time attractors. A similar
situation is seen in case (b) where {\it D1} and {\it C2} are stable.}
\label{num_plots}
\end{figure}

In general, if a cosmological model possesses more than one stable
critical point, amongst the possible late-time solutions it will
reach the one in the basin of attraction of which it lies. Thus,
the specific cosmological evolution depends  heavily  on the
initial conditions. In our scenario, there are regions of the
parameter space where there is both a stable quintessence-like and
a stable phantom-like critical point. Therefore, the EOS can
always remain in the quintessence or the phantom regime, cross the
phantom divide from below to above, or cross the phantom divide
from above to below, with the last possibility mildly favoured by
observations. It must be noted, however, that in the absence of
constraints on the coupling $\alpha$ and the potential exponents,
a late-time solution with $w_{\rm DE}<-1$ seems likely as there is a
stable phantom-like critical point in every region of the
parameter space. This fact is an advantage of the model.

As we know, the standard quintom model \cite{Guo:2004fq,quintom,
Cai:2009zp} exhibits only one (phantom-like) stable late-time
solution, and in order to acquire the simultaneous existence of
one quintessence and one phantom stable late-time solutions, one
has to extend to significantly generalized quintom models
\cite{Setare:2008si}. On the other hand, the model at hand
presents this behavior more economically. Furthermore, note that
the extended scenario of this work exhibits a novel situation,
viz. the presence of more than one stable attractor (points {\it C2} and
{\it D2}) corresponding to quintessence-like solutions. Thus, according
to the specific parameter values, and for suitable initial
conditions, the late-time behaviour of the Universe can be
described by one of the quintessence-like states, with {\it D2}
possessing a more physical EOS compared to {\it C2} (see Table
\ref{table:stability}).

\begin{figure}[ht]
\centering \subfigure[] {
\includegraphics[width=8.6cm]{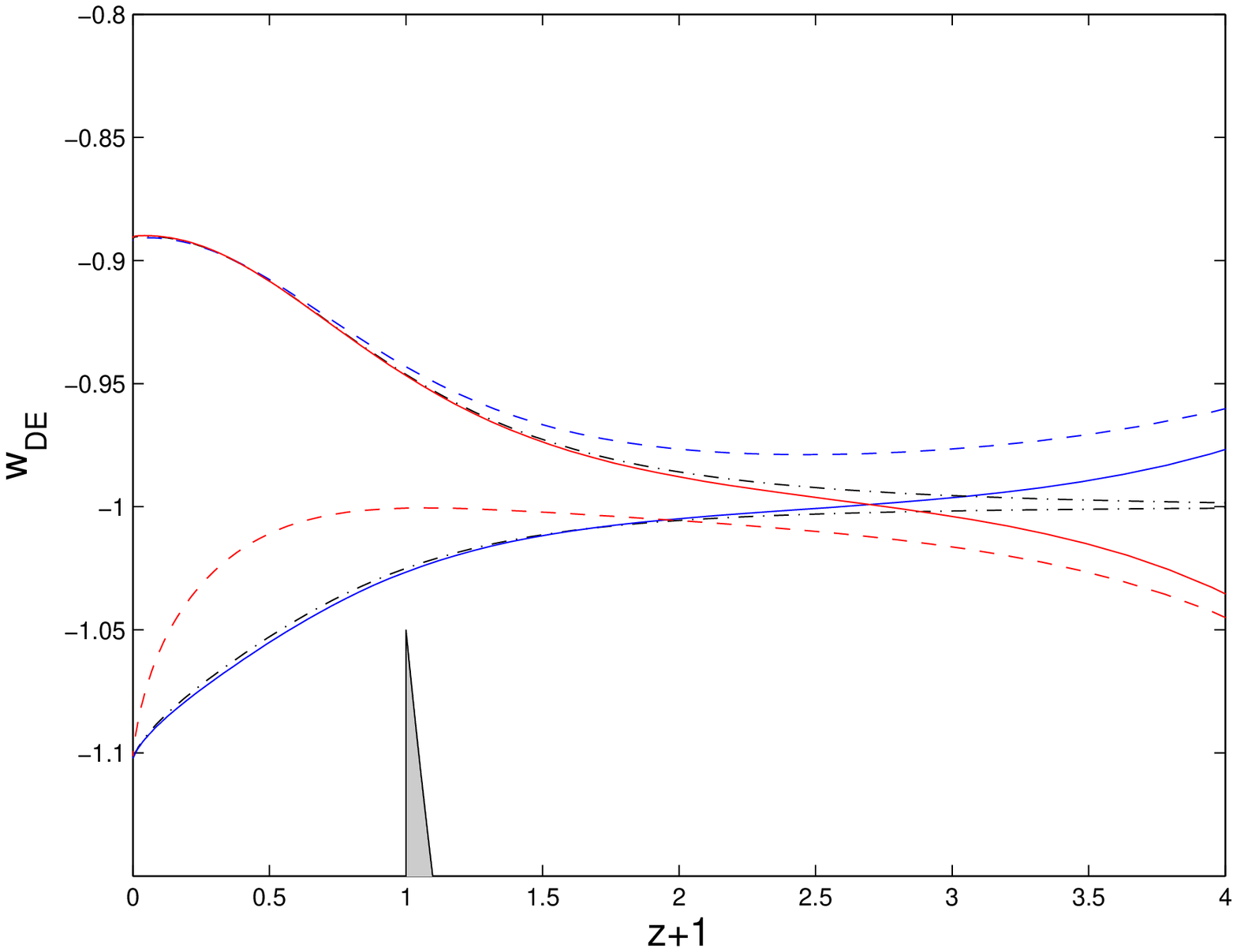}
} \subfigure[] {
\includegraphics[width=8.6cm]{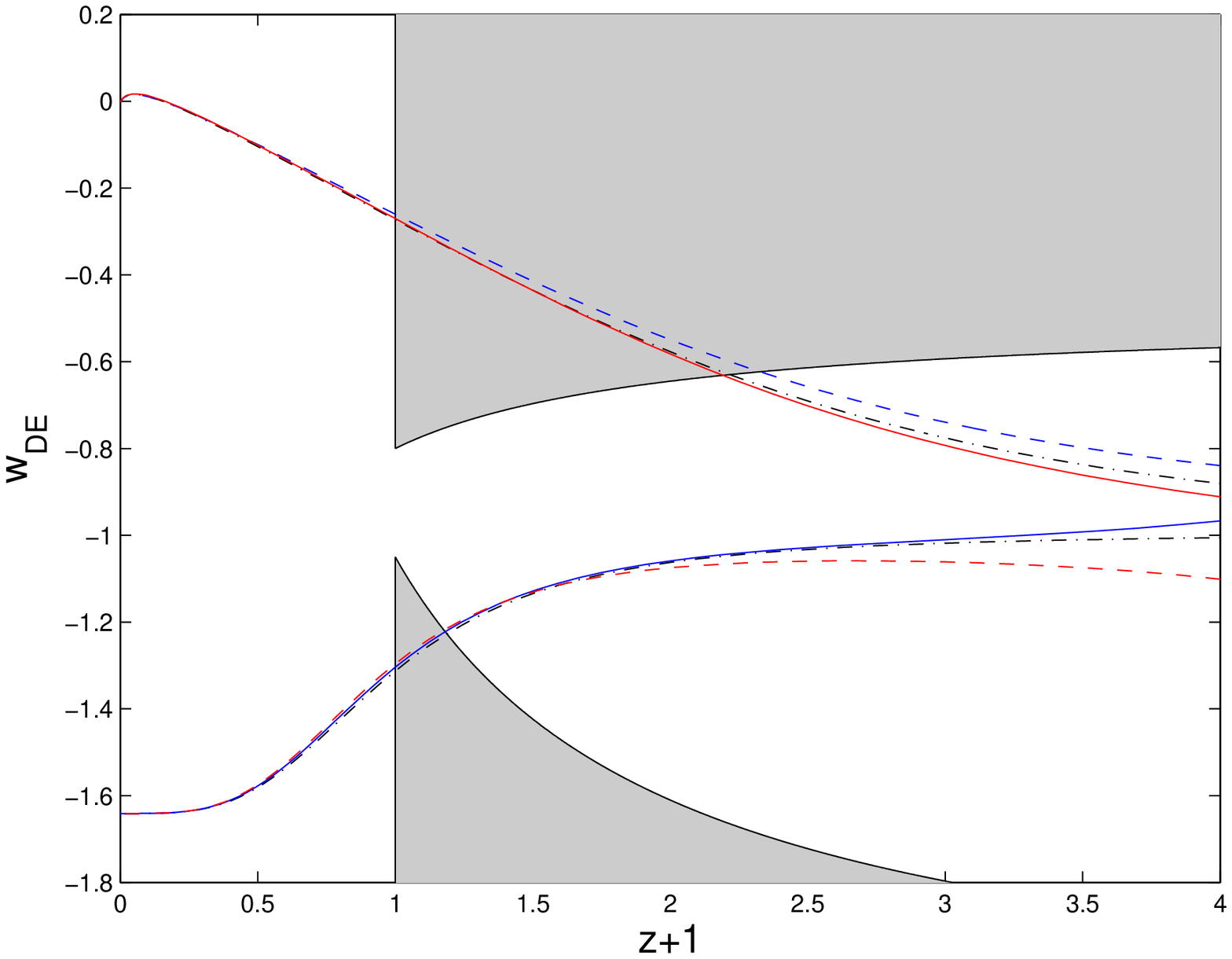}
}
 \caption{
 The evolution of the dark energy EOS with respect to redshift [$z+1 = a(t_0)/a(t)$].
 Parameter values, initial conditions and colour-coding are as in Fig.
 \ref{redshift_plots}. $z=0$ is chosen to correspond to the point at which $\Omega_{DE} = 0.728$.
 Shaded regions are excluded by Wilkinson Microwave Anisotropy Probe 7 yr data+BAO+$H_0$+SN at 68\% confidence level \cite{WMAP_7yrs}.
 Using the parameterisation $w_{\rm DE}=w_0+w_a z/(z+1)$, the limits are $w_0 = 0.93\pm0.13$ and $w_a = -0.41^{+0.72}_{-0.71}$.
 } \label{redshift_plots}
\end{figure}

In order to present  the late-time behaviour transparently, in
Fig. \ref{num_plots} we depict the results of a numerical
integration for two choices of parameter values. In graph (a) the
phantom stable point is {\it D1} and the quintessence one is {\it D2}, while
in graph (b) the phantom stable point is also {\it D1} but the
quintessence one is C2. If the system avoids the stable phantom
solution, it will tend toward one of the two quintessence-like
solutions, depending on the values of the parameters, $\alpha,
\mu$ and $\lambda$.

We have to mention that for a particular subclass of evolutionary
trajectories, depending on the parameter values and initial
conditions, the dark energy EOS becomes infinite at a finite time
$t_*$. However, this divergence in $w_{\rm DE}$ is not accompanied by
a divergence in the scale factor, its time-derivative or in the
dark energy density and pressure, but is an expected result
arising from the nullification of $\rho_{DE}$, possible in all
phantom models. Thus, technically, it does not correspond to the
Big Rip discussed in the literature \cite{phantBigRip}, but rather
to the new singularity family discussed in \cite{Bamba:2008ut}.
One could consider only those models in which the total energy
density remains positive (where the potential is sufficiently
positive or the phantom kinetic energy not too large), to avoid
the corresponding energy condition violation. Finally, note that
while a negative coupling value $\alpha$ facilitates the
nullification of $\rho_{DE}$, a positive $\alpha$ obstructs it.

As one can observe in Fig. \ref{num_plots}, the dark energy EOS
can change significantly before the system finally reaches the
critical points. It is therefore of great interest to investigate
whether the situation described above is
compatible with the stringent bounds provided by cosmological
observations. To this end, in Fig. \ref{redshift_plots} we replot
the evolution of $w_{\rm DE}$ of Fig. \ref{num_plots}, using the
redshift $z$ [$z+1 \equiv a(t_0)/a(t)$] as the independent
variable, presenting also the observationally excluded regions
from \cite{WMAP_7yrs}, up to $z=3$. In the upper graph, both
potentials are relatively flat and the values of $w_{\rm DE}$ at the
critical points {\it D1} and {\it D2} are close to $-1$. This means that,
despite the range of behaviour exhibited for different initial
conditions, the EOS does not deviate much from $-1$ when the
trajectories are approaching the attractors at recent times. These
features make the evolutions of Fig. \ref{redshift_plots}(a)
compatible with observations.

The case in Fig. \ref{redshift_plots}(b) is quite different. As
mentioned previously, it can be seen that the scaling point {\it C2} is
ruled out as a late-time solution. Also, for both {\it C2} and {\it D1} to be
stable, $\lambda$ needs to be larger, leading to a more negative
EOS for {\it D1}. As the trajectories start to converge on this value,
$w_{\rm DE}$ evolves more quickly than Fig. \ref{redshift_plots}(a),
leading to an EOS that is disfavoured by observations at $z\approx
0$.

The behaviour of  Fig. \ref{redshift_plots} is fairly typical. In
the case where the dark energy contribution to the energy content
of the Universe is initially negligible, the value
$\Omega_{DE}\approx 0.7$ is obtained only when the solutions start
to converge to the critical points. Thus, if the value of $w_{\rm DE}$
at the critical points is closer to $-1$, the variation in the EOS
is smaller at recent times, and closer to the observational
values.

In order to perform a complete comparison with the data, it is
necessary to take into account the effect of dark energy
perturbations. In particular, the late-time integrated Sachs-Wolfe
 effect, which affects the CMB on large angular scales, is
sensitive to the presence of dark energy perturbations
\cite{DEreview}. Perturbations in the standard quintom model have
been investigated in \cite{Zhao:2005vj}, where it was found that
the parameter space that allowed the EOS to cross the phantom
divide was enlarged when the perturbations were included. As the
present scenario exhibits behaviour similar to simple quintessence
and quintom models, we expect that many features of these
scenarios will be retained when the perturbations are taken into
account. However, as we have seen, an important difference between
the coupled and uncoupled cases is the presence of a source term
proportional to $\alpha$ that can drag the quintessence (phantom)
field up (down) its potential. This would introduce extra terms
into the perturbation equations which could lead to observable
effects. Another interesting possibility is that the behaviour of
isocurvature perturbations will be affected by the transfer of
energy between the two fields. The full analysis of the
perturbations in the model at hand is beyond the scope of the
current investigation and will be treated in a future work
\cite{us.future}.

\section{Conclusions}
\label{conclusions}

We have considered an extension of the quintom model of dark
energy in which the kinetic terms of the phantom  and the
canonical scalar field are coupled by a term in the Lagrangian of
the form  $\tfrac{\alpha}{2} g^{\mu\nu}\p_\mu\phi\p_\nu\sigma$. In
order to study the asymptotic behaviour of the model and to
facilitate comparison both with the standard quintom scenario and
the two-field quintessence model with mixed kinetic terms, we have
performed a phase-space analysis of the corresponding dynamical
system.

We find that the kinetic interaction allows for the possibility of
stable critical points similar to those found in quintessence
scenarios, including field-dominated solutions with $w_{\rm DE}>-1$
and solutions displaying scaling behaviour. Additionally, there
exist two new critical points {\it E} and {\it F} that are not present in
the simple phantom model, the latter of which is analogous to that
responsible for the phenomenon of assisted quintessence in the
case with two canonical fields. Here, however, the combined effect
of the fields is to give phantom-like behaviour, with $w_{\rm DE}<-1$.
As well as this, there is an extension of the phantom-dominated
asymptotic solution in the standard quintom model, which is stable
for a large region of the parameter space. In all, there is a
stable solution with $w_{\rm DE}<-1$  and $\Om=1$ in every part of the
parameter space, even when the quintessence-like points are
stable. We have discussed the cosmological implications of these
results in Sec. \ref{cosmimpl}.

A characteristic feature of the presence of the kinetic coupling
is the presence of stable solutions in which the quintessence
field can be dragged up its potential, in the opposite direction
to the standard case. This occurs when the coupling is negative,
where the quintessence equation of motion acquires a positive
source term. Similarly, for a positive coupling the phantom-field
equation of motion obtains a negative source term, which causes
the field to roll down its potential. This could affect the
evolution of the dark energy perturbations and lead to
observational differences between the kinetically coupled quintom
scenario and other models, offering a relatively safe signature.

The use of canonical and phantom fields to drive the acceleration
of the Universe in phenomenological models is not necessarily
representative of the fundamental mechanism behind dark energy, in
which non-trivial features such as non-canonical kinetic terms and
kinetic couplings may be important. The scenario of the present
work is interesting as the kinetic coupling gives rise to
behaviour different from that of the standard quintom model. In
particular, both quintessence-like and phantom-like late-time
solutions, as well as solutions that cross the phantom divide
$w_{\rm DE}=-1$  during the evolution, are possible. Thus, such
scenarios could be a candidate for the description of dark energy.

\begin{acknowledgments}

We would like to thank Carsten van de Bruck for useful discussions
and for comments on the manuscript. JW is supported by EPSRC.

\end{acknowledgments}

\appendix*

\section{The linearised perturbation matrix}
\label{appendix}

The $4\times4$ matrix ${\bf {Q}}$ of the linearised perturbation
equations of the system (\ref{eq:Xia}) reads:
\[{\bf {Q}}= \left[ \begin{array}{cccc}
Q_{11} & Q_{12}& Q_{13} &Q_{14}  \\
Q_{21} & Q_{22}& Q_{23} &Q_{24}  \\
Q_{31} & Q_{32}& Q_{33} &Q_{34}  \\
Q_{41} & Q_{42}& Q_{43} &Q_{44}
\end{array} \right],\]
with
\begin{eqnarray}
Q_{11}&=&  -3+\tfrac{3}{2}x_{\phi c}(2-\gamma)(2x_{\phi c}+\alpha x_{\sigma c})+T_c,       \ \  \nn\\
Q_{12}&=& \tfrac{3}{2}x_{\phi c}(2-\gamma)(-2x_{\sigma c}+\alpha x_{\phi c}), \nn\\
Q_{13}&=& y_{\phi c}\left[
\frac{4\sqrt{6}\lambda}{(4+\alpha^2)}-3\gamma x_{\phi c}
\right],\nn\\
Q_{14}&=& y_{\sigma c}\left[
\frac{2\alpha\sqrt{6}\mu}{(4+\alpha^2)}-3\gamma x_{\phi c} \right],
\nn
\end{eqnarray}
\begin{eqnarray}
Q_{21}&=&  \tfrac{3}{2}x_{\sigma c}(2-\gamma)(2x_{\phi c}+\alpha x_{\sigma c}), \nn\\
Q_{22}&=& -3+\tfrac{3}{2}x_{\sigma c}(2-\gamma)(-2x_{\sigma c}+\alpha x_{\phi c})+T_c,  \nn\\
Q_{23}&=& y_{\phi c}\left[ \frac{2\alpha\sqrt{6}\lambda}{(4+\alpha^2)}-3\gamma x_{\sigma c} \right], \nn\\
Q_{24}&=&-y_{\sigma c}\left[
\frac{4\sqrt{6}\mu}{(4+\alpha^2)}+3\gamma x_{\sigma c} \right], \nn
\end{eqnarray}
\begin{eqnarray}
Q_{31}&=& y_{\phi c}\left[ \tfrac{3}{2}(2-\gamma)(2x_{\phi c}+\alpha x_{\sigma c})-\sqrt{\tfrac{3}{2}}\lambda\right], \ \  \nn\\
Q_{32}&=& \tfrac{3}{2} y_{\phi c}(2-\gamma)(-2x_{\sigma c}+\alpha
x_{\phi c}), \nn
\end{eqnarray}
\begin{eqnarray}
Q_{33}&=& -\sqrt{\tfrac{3}{2}}\lambda x_{\phi c}-3\gamma y_{\phi c}^2 +T_c,\ \ \ \ \ \ \ \ \ \ \ \ \ \ \ \ \ \ \ \  \nn\\
Q_{34}&=& -3\gamma y_{\phi c} y_{\sigma c}, \nn
\end{eqnarray}
\begin{eqnarray}
Q_{41}&=&  \tfrac{3}{2} y_{\sigma c}(2-\gamma)(2x_{\phi c}+\alpha x_{\sigma c}),  \nn\\
Q_{42}&=& y_{\sigma c}\left[ \tfrac{3}{2}(2-\gamma)(-2x_{\sigma c}+\alpha x_{\phi c})-\sqrt{\tfrac{3}{2}}\mu\right],  \nn\\
Q_{43}&=&  -3\gamma y_{\phi c} y_{\sigma c}, \nn\\
Q_{44}&=&  -\sqrt{\tfrac{3}{2}}\mu x_{\sigma c}-3\gamma y_{\sigma
c}^2 +T_c,
\end{eqnarray}
where
\begin{eqnarray}
 T_c&=&\frac{3}{2}\Big[  2 x_{\phi c}^2 -2 x_{\sigma c}^2+2\alpha x_{\phi c}
x_{\sigma c}
+ \nonumber\\
&\ &+\gamma (1-x_{\phi c}^2+x_{\sigma c}^2-\alpha x_{\phi c}
x_{\sigma c}-y_{\phi c}^2-y_{\sigma c}^2)\Big ].\ \ \ \ \ \
\label{eq:Tc}
\end{eqnarray}

\end{document}